\documentclass[twocolumn,pre,showpacs,preprintnumbers,amsmath,amssymb,superscriptaddress]{revtex4}

\usepackage[english]{babel}
\usepackage{graphicx}
\usepackage{dcolumn}
\usepackage[colorlinks=false]{hyperref}
\usepackage{bm}
\usepackage{subfigure}
\usepackage{longtable}

\graphicspath{
    {figures/}
}

\def\levy{L\'evy }

\begin{document}
\preprint{-}
\title{Superdiffusion and Transport  in $2d$-systems with \levy Like Quenched Disorder}
\author{}
\affiliation{}

\author{Raffaella Burioni}
\affiliation{Dipartimento di Fisica e Scienza della Terra, Universit\`a di
Parma, viale G.P.Usberti 7/A, 43124 Parma, Italy}
\affiliation{INFN, Gruppo Collegato di
Parma, viale G.P. Usberti 7/A, 43124 Parma, Italy}
\author{Enrico Ubaldi}
\affiliation{Dipartimento di Fisica e Scienza della Terra, Universit\`a di
Parma, viale G.P.Usberti 7/A, 43124 Parma, Italy}
\author{Alessandro Vezzani}
\affiliation{Centro S3, CNR-Istituto di Nanoscienze, Via Campi 213A, 41125 Modena Italy}
\affiliation{Dipartimento di Fisica e Scienza della Terra, Universit\`a di
Parma, viale G.P.Usberti 7/A, 43124 Parma, Italy}


\date{January 2013}

\begin{abstract}
We present an extensive analysis of transport properties in superdiffusive two
dimensional quenched random media, obtained by packing disks with radii
distributed according to a L\'evy law. We consider transport and scaling
properties in samples packed with two different procedures, at fixed filling
fraction and at self-similar packing, and we clarify the role of the two
procedures in the superdiffusive effects. Using the behavior of the filling
fraction in finite size systems as the main geometrical parameter,
we define an effective L\'evy exponents that correctly estimate the finite size
effects. The effective L\'evy exponent rules the dynamical scaling of the main
transport properties and identify the region where superdiffusive effects can be
detected.
\end{abstract}

\maketitle{}

\section{Introduction}

Transport  and diffusion in highly heterogeneous random media is an interesting
and complex phenomenon occurring in a wide class of materials, ranging from
rocks to clouds, to engineered experimental samples
\cite{PhysRevLett.65.2201,Davis:2002ys,Benson:2001zr,Palombo:2011mz,Brockmann:2003uq,PhysRevLett.72.203,PhysRevLett.54.616}.
Matter or light typically crosses disordered regions with very different
diffusion properties, so that  the transport process can be described as a
L\'evy walk: motion consists of a sequence of scatterings  followed by long
jumps in the nonscattering regions. Two concurrent effects make the problem
particularly hard: the  quenched randomness, inducing correlation between
displacements, and,  if the material is heterogeneous on all scales,  the
typical broad distribution of the steps length, that can be heavy-tailed and
L\'evy distributed.

Tunable media with L\'evy like properties \cite{Barthelemy:2008rt}, reproducing
all these effects, are an interesting testbed for the typically superdiffusive
motion occurring in heterogeneous random materials. The "L\'evy" glasses are
obtained by a packing of polydispersed glass spheres with L\'evy distributed
radii and embedded in a scattering matrix. As light is scattered only in the
inter-sphere regions and freely travels ballistically inside the spheres, the
propagation of light in these materials can be described as a  correlated L\'evy
walk, with step length distribution $p(l)\sim l^{-(1+\alpha)}$, where the
parameter $\alpha$ can be tuned by choosing an appropriate distribution of radii
for the polydispersed spheres. In these systems, a superdiffusive behavior has
been observed, with the characteristic length of the dynamical process growing as
$\ell(t)\sim t^{1/z}$ with $z<2$.

A theoretical study of such systems is in general a non trivial task 
due to topological correlations between the step lengths: e.g. after crossing
a large sphere a walker is likely backscattered in the same ballistic region.
In this perspective, different theoretical and numerical approaches
have been proposed \cite{barthelemy:2010PRE,beenakker:2012,svenson:2013PRE}.
An analytic solution has been obtained for the simpler one dimensional case, the 
so called L\'evy-Lorentz gas, 
corresponding to a packing of segments of  L\'evy  distributed lengths separated by 
scattering centers \cite{PhysRevE.61.1164}; in that case
the dynamical exponent is  $z=1+\alpha$ (superdiffusion) for $0<\alpha<1$ and
$z=2$ (standard diffusion) for $\alpha>1$ 
\cite{Beenakker:2009ys,Burioni:2010qy,Burioni:2010fk,Disanto}.
The regular and deterministic self similar version of the L\'evy packings in
higher dimension \cite{Buonsante:2011} have been recently introduced as basic
geometrical models for quenched L\'evy structures. Also in that case, numerical
simulations show superdiffusion for $\alpha<1$ and  standard diffusion for
$\alpha>1$, with the dynamical exponent $z$ depending on $\alpha$,  and very
slightly on the dynamical rule and on the spatial dimension. Conversely, recent
numerical results on $2$ and $3-$dimensional random sphere packings
\cite{beenakker:2012} evidenced important differences with previous results. In
particular superdiffusion was observed also for $\alpha>1$, i.e. for $\alpha
\lesssim 1.6$. However, some aspects remained unclear, in particular the
anomalous non monotonic behavior of $z$ as a function of $\alpha$, and the
evaluation of the finite size effects \cite{beenakker:2012}. 
Notice that, for these random high dimensional heterogeneous models, even the
definition of the packing procedure is a complex problem, still not understood
from the theoretical point of view
\cite{Biazzo:2009qy,Parisi:2010,Torquato:2010}.
What seems clear is that the dynamical behavior observed in all these L\'evy quenched structures
differs from the standard uncorrelated L\'evy walk, where anomalous
superdiffusion is present for $1<\alpha<2$ \cite{ann1,ann2,ann3}.

Another important point in the numerical experiments performed so far is that 
the samples are built by optimizing the packing, and minimizing the scattering region 
(optimal filling approach). In this way, the average distance 
between the spheres is kept constant as the system size grows, reproducing 
the behavior of deterministic fractals \cite{Buonsante:2011}.
Within this prescription, the density of scatterers (or turbid fraction), 
for $\alpha<1$, becomes infinitesimal  as the size goes to 
infinity, as one is building a so called slim fractal.
Conversely, in the samples produced for the experiments \cite{Barthelemy:2008rt}
the turbid fraction is kept constant at different sizes even for $\alpha <1$, 
and this may give rise to important differences with respect the optimized L\'evy structures of the
theoretical approaches.

In this paper we first present an extensive numerical and theoretical analysis of
two dimensional random packings of disks with radii distributed according
to a L\'evy law, and packed by minimizing the turbid fraction, i.e. with the
optimal filling approach. Our results provide a new interpretation of the data
found in  \cite{beenakker:2012}. In particular, the anomalous non monotonic
behavior of the dynamical exponent $z$ can be interpreted as a failure of the
packing procedure in the regime $\alpha<0.5$. Moreover, numerical simulations
show that  in the thermodynamic limit of infinitely large samples the
superdiffusive regimes observed for $1<\alpha \lesssim 1.6$ are expected to disappear and
converge to a behavior similar to the deterministic case, where $z<2$ only for
$\alpha<1$. In this framework we evidence that for $0.4 \lesssim \alpha \lesssim
1.6$ strong finite size effects characterize the dynamics, the
radii distribution and the packing procedure. We define the exponent
$\alpha_{eff}$, describing such geometrical preasymptotic behavior, and we show
that  $\alpha_{eff}$ can be used for an effective description of the size
effects characterizing the dynamics of realistic finite samples. Finally, we
discuss the effect on the dynamics of the scattering length and of the
truncation of the radii distribution.

Then, in order  to make contact with the experiments, we consider the
case of disks packing obtained by  keeping the filling fraction constant at all scales,
instead of minimizing the turbid fraction. This analysis evidences that, for any
value of $\alpha$, as the size of the systems grows, the dynamical exponent $z$
becomes closer to 2, and we infer that the system is experiencing a crossing
from a superdiffusive to a diffusive regime, with a large crossover region where
superdiffusive effects can appear. The evidence is corroborated by analyzing the
response of the dynamical exponents to the change of the scattering mean free
path length and the fixed filled fraction $f$. We remark that such a crossover 
from ballistic to diffusive behavior is also present in the case where all the spheres
have the same radius \cite{singler}.

The outline of the work is as follows. In Section \ref{sec:model} we introduce
the dynamics and the optimized packing procedure. In Section \ref{sec:ff} we
study the geometrical properties of the samples, introducing the effective
exponent. We then show the results of our dynamical simulations and scaling
effects in Section \ref{sec:dynamics}. Finally, Section \ref{sec:experiment} is
devoted to the simulations at fixed filling, designed to make contact with the
experiments. Section \ref{sec:conc} contains our conclusions.

\section{Random  L\'evy packings and dynamics}
\label{sec:model}

The packing algorithm follows the work of Beenakker et al. \cite{beenakker:2012} and the parameters of the random L\'evy structure are:
\begin{itemize}
    \item [-] $N$, the number of disks (spheres);
    \item [-] $L_x,\,L_y$ the thickness and the width of the slab, respectively;
    \item [-] $r_{min}$ and $r_{max}$,  the minimum and the maximum radii, respectively;
    \item [-] $\beta$, the exponent used to generate the radii distribution,
        featuring a power law  $p(r)\sim r^{-(1+\beta)}$. We set $\alpha=
        \beta - d +1$, with $d=2$ in the case of disks.
\end{itemize}

The latter distribution is sampled following the recursive formula  \cite{beenakker:2012}:
\begin{eqnarray}
r_{k}={}&r_{\rm max}\left[1+\frac{k}{k_{\rm max}}(r_{\rm max}^{\beta}-1)\right]^{-1/\beta},\nonumber\\
&\;\;k=0,1,2,\ldots k_{\rm max},
\label{eq:rkdef}
\end{eqnarray}
generating the radii ordered by size from the largest to the smallest one.
According to \cite{beenakker:2012} disks can be placed randomly in the systems
avoiding overlaps. 
The filling fraction $f$ at size $L$ is defined as the ratio between the total volume of the
disks placed in the sample, $V_D(L)$, and the total volume $V(L)$ of the sample:
\begin{equation}
    f(L) = \frac{V_D(L)}{V(L)}.
    \label{eq:ff}
\end{equation}
The complementary of $f$ is the turbid fraction $\phi(L)=1-f(L)$, that is, the
system volume fraction occupied by the scatterers.

\begin{figure}
    \centering
    {\includegraphics[width=.9\columnwidth]{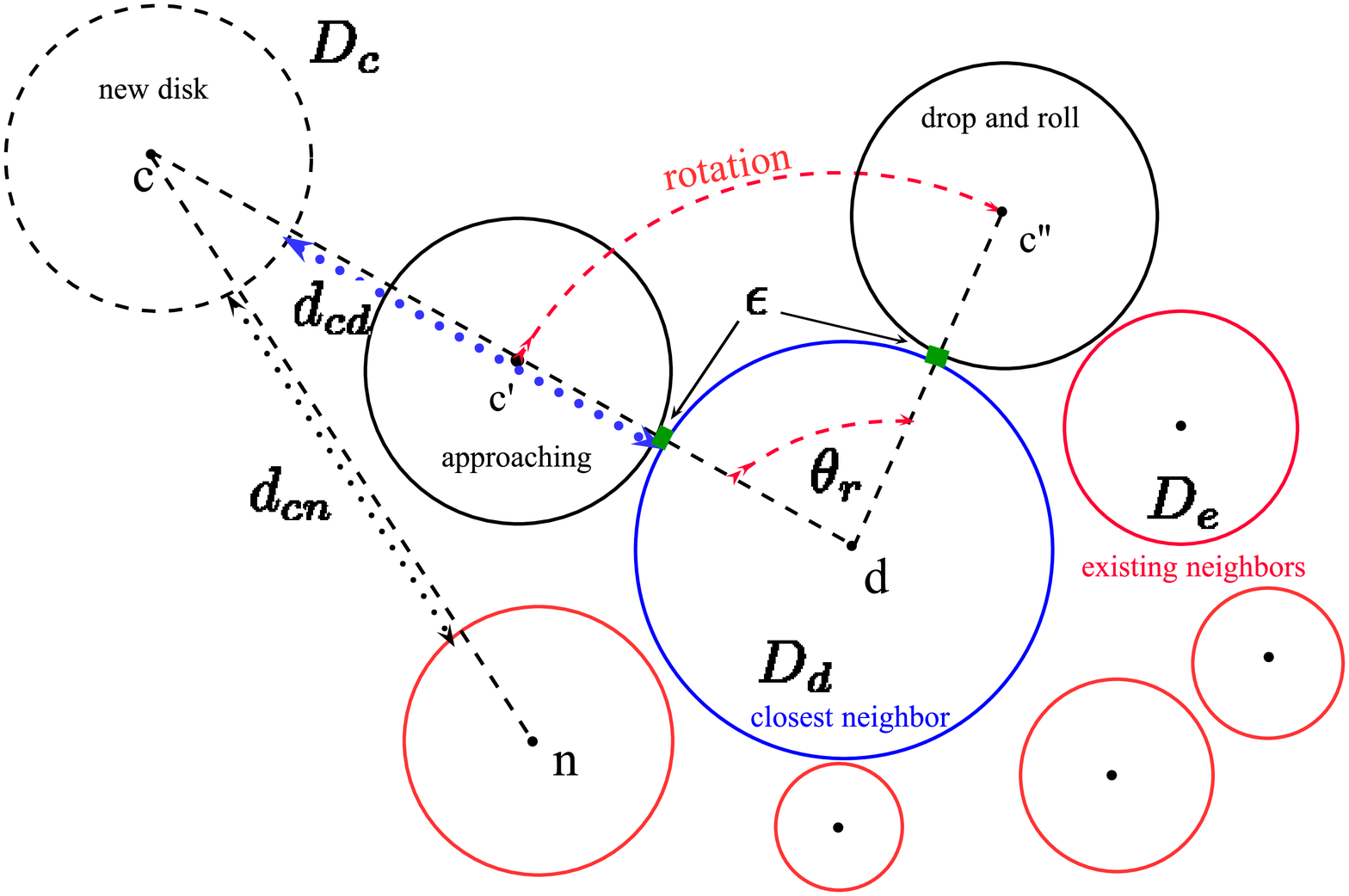}}
    \newline
    {\includegraphics[width=.9\columnwidth]{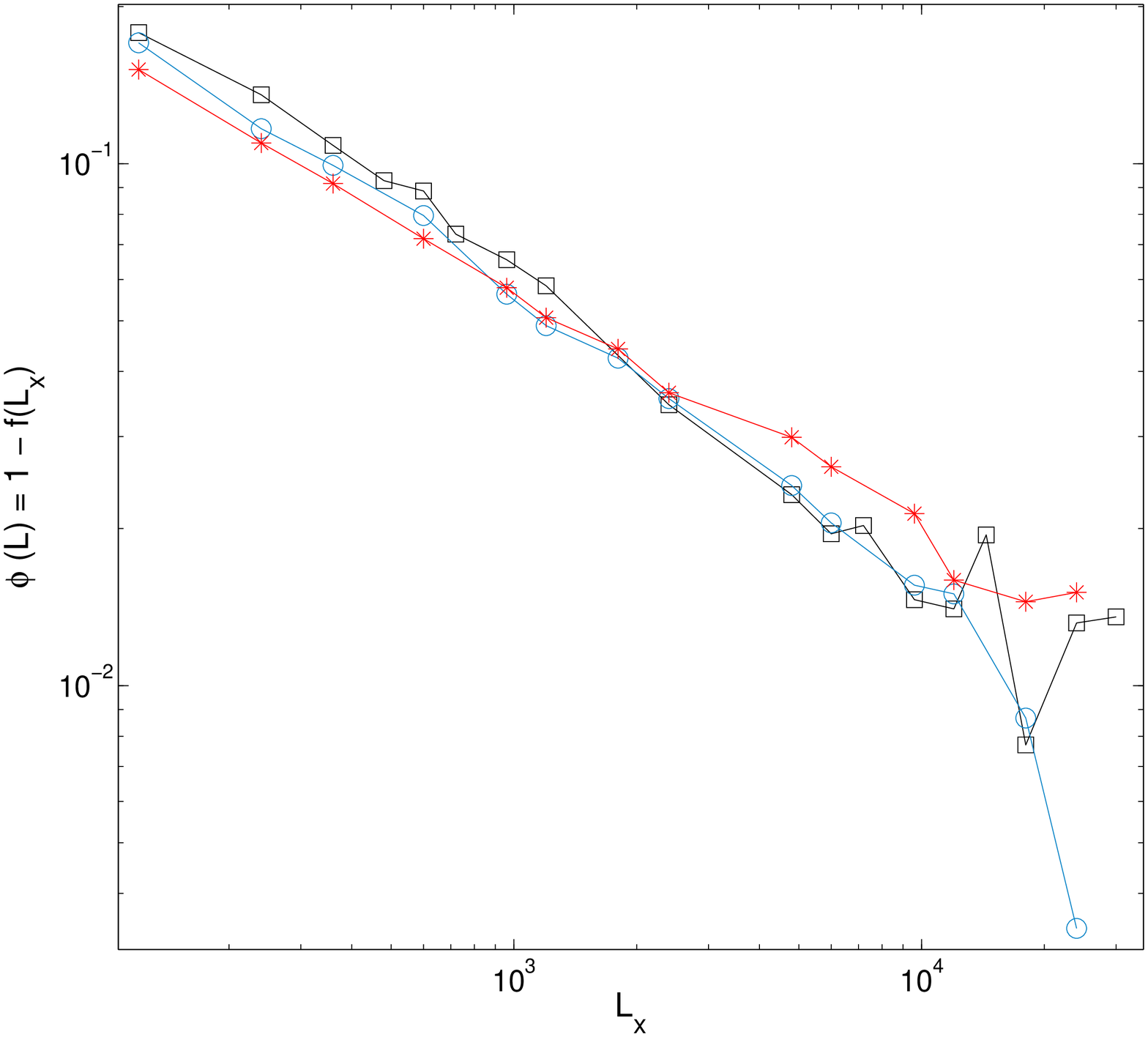}}
    \caption{
        (Color online) (Top Panel) The geometric outline of the two methods implemented. When a
        new disk with center $c$ is placed, the program searches for the closest
        neighbor among the close disks (the disk with the $d$ center in this
        case). The distance between them is computed and the disk moved to a
        $\epsilon$ distance placing it in $c'$ (close packing prescription,
        $\epsilon$ randomly chosen from a flat distribution in the $(10^{-4},
        1/4)\cdot r_{min}$).
        In the drop and roll method we further move the new disk rotating it
        around the closest neighbor until it touches an existing disk close to
        it.
        (Bottom Panel) Turbid fraction from the simulations at optimized
        filling following the three prescription presented: random packing
        (squares), dense packing prescription (circles) and the drop and roll
        algorithm (asterisks). The curves correspond to  $\alpha  = 0.4$.
    }
    \label{fig:drop_n_roll}
\end{figure}

\begin{figure}
    \centering
    \subfigure[]
        {\includegraphics[width=.9\columnwidth]{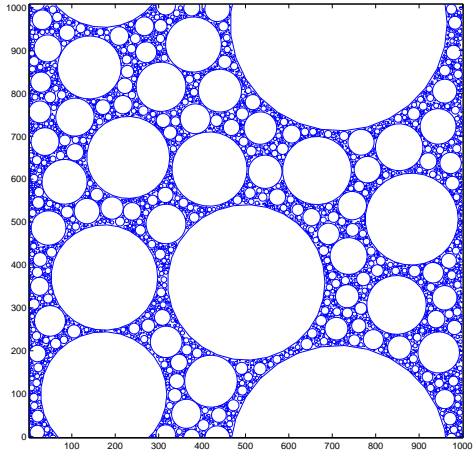}}
        \newline
    \subfigure[]
        {\includegraphics[width=1\columnwidth]{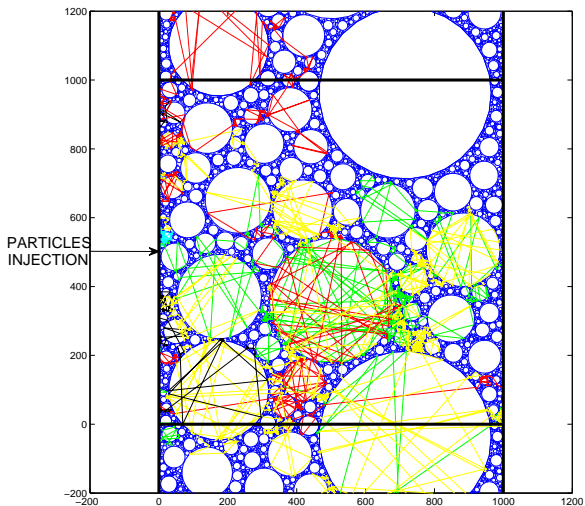}}
        \caption{
            (Color online) (a) A \levy glass with $L_x=L_y=10^3$, $r_{max}=250$, $\alpha=0.4$
            and $N_{disk}=4\cdot10^3$. The geometry is constrained on the $x=0$
            and $x=L_x$ boundaries, while the $y=[0, L_y]$ ones are periodic;
            (b) The dynamics on the same sample. The program injects particles
            into the $x=0$ surface and let them diffuse until they reach the
            opposite side of the slab or they are backscattered to $x=0$. The
            original cell is framed with solid black line. The periodic bounds
            allow the motion to continue over the $y=0$ and $y=L_y$ borders.
        }
    \label{fig:sistema}
\end{figure}

We introduce two different packing
procedures illustrated in the Top Panel of Fig. \ref{fig:drop_n_roll}. In the
first, the newly added disk is  approached to its closest neighbor, in the
second it is also rotated around the neighboring disk until it is still not
overlapping with the rest of the systems \cite{0953-8984-16-37-002,Packing_Xu}.
The results of this new procedures are reported in Fig. \ref{fig:drop_n_roll}
(bottom panel). The data clearly show that both prescriptions give the same
qualitative result, presenting only a small improvement in the filling fraction.
Therefore, hereafter, we adopt the original procedure which is the less
demanding from a computational point of view. An example of a \levy disks
packing produced by this algorithm is found in Fig. \ref{fig:sistema}.

We then implement the specific dynamics of transport in the \levy packing.
The rays of light experience a ballistic motion inside the disks and an
isotropic Poisson process in between the spheres, i.e. in the turbid region.

\begin{figure}
    \centering
    \includegraphics[width=.8\columnwidth]{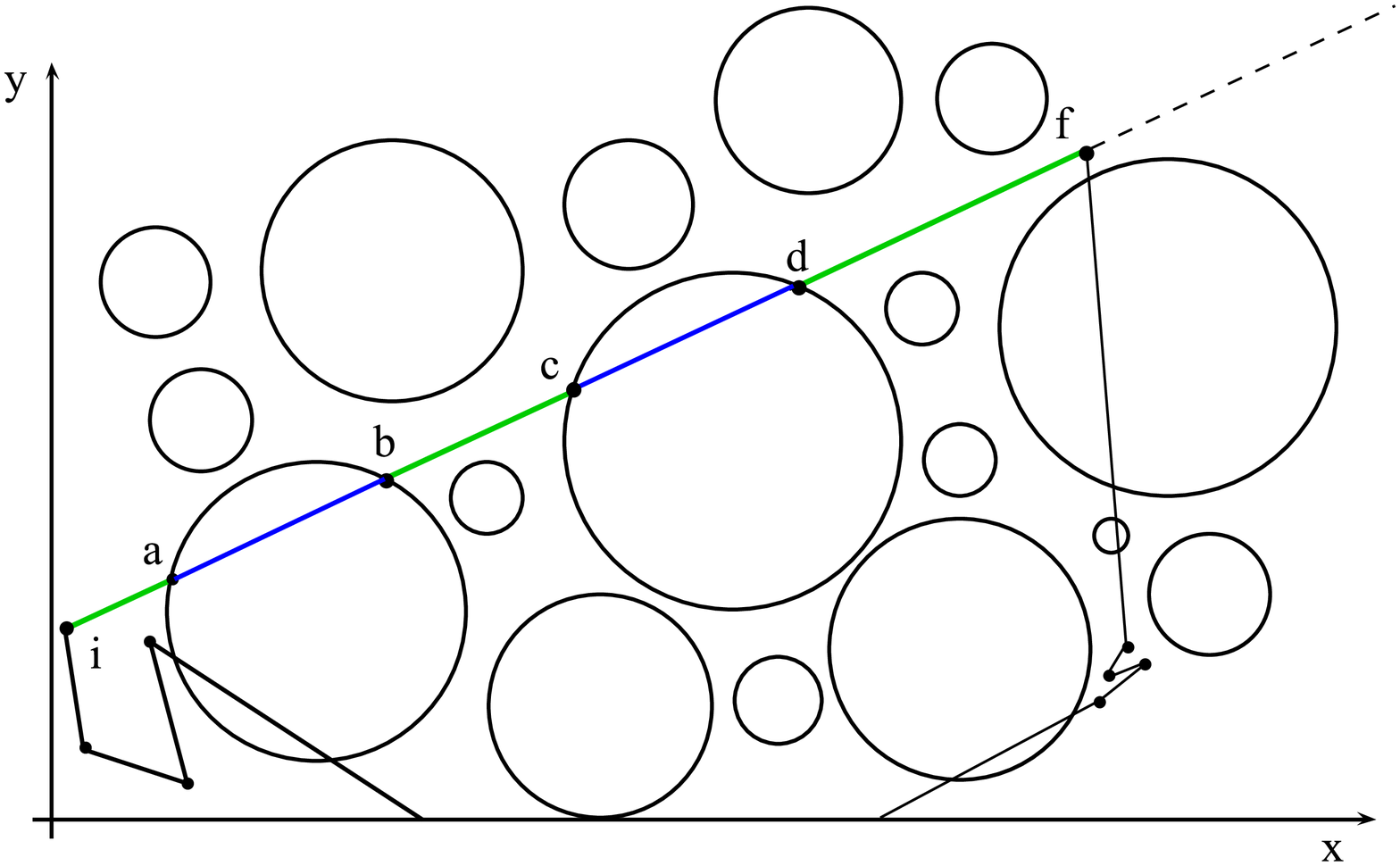}
    \caption{
        (Color online) An example of a trajectory: the single step length $s$ extracted
        from Eq. (\ref{eq:step_length_distrib_alg}) is the length that the ray
        covers outside the disks between two scattering events. The distance
        covered within them is not taken into account.
        In this figure the extracted length is $s = \bar{ia} + \bar{bc} +
        \bar{df}$, while the actual step length is $l = \bar{if}$ due to the
        time spent by the ray within the disks.
        The latter is $\bar{s} = \bar{ab} + \bar{cd}$, so that $l = s +
        \bar s$.
    }
    \label{fig:single_step}
\end{figure}

In particular, at each move we extract a random direction and a random step
length $s$ from the distribution
\begin{equation}
    P(s)ds=\frac{1}{\lambda}\exp{(-s/\lambda)}ds,
    \label{eq:step_length_distrib_alg}
\end{equation}
being $\lambda$ the scattering mean free path. If the ray crosses one or more
disks, the actual step length has to be incremented until the ending point
belongs to the turbid region and the new step length becomes $l=s+\bar{s}$
($\bar{s}$ is the spaced covered within the disks). An example of a trajectory
is shown in Fig. \ref{fig:single_step}.

In the transmission simulations we select  the initial point on the $x=0$
side of the slab. The particles then walk as long as they reach the opposite
side of the slab ($x=L_x$) or until they are backscattered to the $x=0$ surface
(see Fig. \ref{fig:sistema}). For the measure of the
probability distribution $P(r,t)$ and of the mean square displacement,
we pack the disks with periodic boundary conditions and the starting
point is chosen randomly in the turbid fraction.

We let the maximum radius range from $r_{max}=10$ to
$r_{max}=1500$, with $L_x=R_S r_{max}$ ($R_S=6$) and $L_y=4L_x$.
The number of disks ranges from few hundreds at small system sizes
to $\sim 6\cdot10^7$.
We initially choose $\lambda=r_{min}=1$.
For each transmission run we simulate $5\cdot10^6$ rays, while in the mean
square displacement and $P(r,t)$ runs only $10^5$ rays are simulated, due to the
time-consuming computation of the single ray trajectory.

\section{The filling fraction and the effective $\alpha$ exponent}
\label{sec:ff}

\begin{figure}
    \centering
    \includegraphics[width=.9\columnwidth]{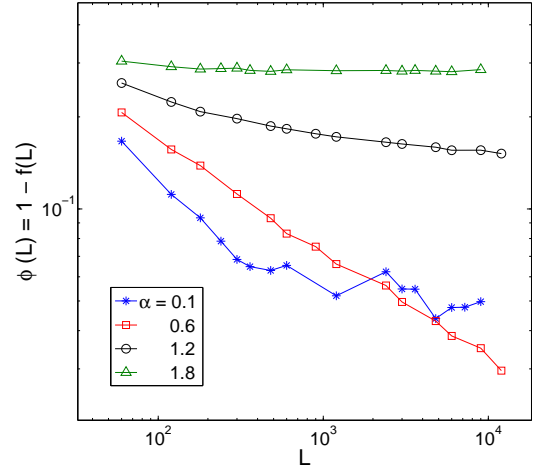}
    \caption{
        (Color online) The numerical behavior of $\phi(L)$ as a function of $L$ for
        $\alpha=0.1$ (asterisks), $\alpha=0.6$ (squares), $\alpha=1.2$
        (circles), $\alpha=1.8$ (triangles).
    }
    \label{fig:ff}
\end{figure}

Let us now consider the behavior of the filling and turbid fraction. Figure
\ref{fig:ff} describes, in our packing, the turbid
fraction as a function of $L$. In large $L\to \infty$ systems we evidence 
that: for $\alpha\lesssim 0.5$, $\phi(L)$ goes to a constant; in
the intermediate regimes $0.5\lesssim \alpha \lesssim 1.6$, $\phi(L)$ vanishes
at large size with a characteristic exponent that we call $\alpha_{exp}$, i.e.
$\phi(L)\sim L^{\alpha_{exp}-1}$; finally for $\alpha\gtrsim 1.6$, $\phi(L)$
goes to a constant in the large $L$ limit.
We remark that the behavior of $\phi(L)$ is here more complex than that observed
in deterministic self similar L\'evy structures\cite{Buonsante:2011}. In that
case, the packing is regular, exact and it is built by a recursive procedure.
There, $\phi(L)\sim L^{\alpha-1}$ for $\alpha<1$ and $\phi(L)$ goes to a
constant for $\alpha>1$, defining the so called slim and fat fractals,
respectively.

\begin{figure}
    {\includegraphics[width=.95\columnwidth]{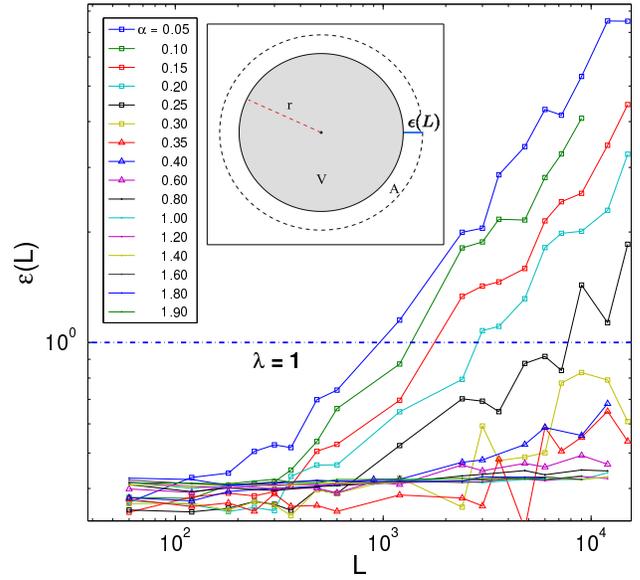}}
    \caption{
        (Color online) Numerical data evidence that $\epsilon(L)$ depends on the
        system size $L$.
        Each line refers to a different $\alpha$ in the $0.1\le\alpha\le1.8$
        range which is displayed from top (dark grey squares $\alpha= 0.1$)  to bottom
        (squares, triangles and dots).  The function diverges only for $\alpha\lesssim
        0.5$,  while at higher $\alpha$ it is constant. The horizontal dashed
        line denoted the scattering length $\lambda = 1$.
        (Inset) On average, around each sphere there is an empty region
        whose thickness is defined to be $\epsilon$, the filling fraction $f$ is
        related to $\epsilon$ and the average volume $V$ and surface $A$
        according to formula (\ref{eq:eps}).
}
        \label{fig:epsilon_L}
\end{figure}

Another  useful quantity for the description of the packing properties is the
average distance $\epsilon$ between two spheres (see Fig. \ref{fig:epsilon_L}
inset).  We assume that, on the average, around each sphere there is a turbid
region whose thickness $\epsilon(L)$ can depend on the system size $L$.  This
quantity is related to the turbid fraction by:
\begin{equation}
    1-f(L) = \frac{\epsilon(L) A(L)}{V(L)} f(L)
    \label{eq:eps}
\end{equation}
where $A(L)$ is the average surface of the spheres.

The ratio $A/V$ as a function of the systems size is  independent 
of the packing algorithm, and can be evaluated directly from the sphere 
distribution Eq. (\ref{eq:rkdef}). In particular, if $k_{max}$ is large
enough, the effect of discretization is negligible, and we can set $p(r)\sim r^{-(1+\beta)}$. We have therefore:
\begin{eqnarray}
    A_c(d,r_{max}) &=& C_A(d)\int_{1}^{r_{max}}{r^{d-1}p(r)dr} \nonumber \\
    V_c(d,r_{max}) &=& C_V(d)\int_{1}^{r_{max}}{r^{d}p(r)dr}.
    \label{eq:v_a_continui}
\end{eqnarray}
where $C_A(d)$ and $C_V(d)$ are the constants defining the surface and the
volume, respectively, at a given dimension $d$ (for instance $C_A(2) = 2\pi$ and
$C_V(3) = 4/3 \pi$).  We finally obtain:
\begin{equation}
    \frac{V}{A}\left(r_{max}\right) = \frac{C_V(d)}{C_A(d)} \frac{r_{max}^{-\alpha + 1} - 1}{r_{max}^{-\alpha} - 1} \frac{-\alpha}{-\alpha + 1}.
    \label{eq:analyt_finite}
\end{equation}
From this equation, we find in the thermodynamic limit
\begin{equation}
    \frac{V}{A}\left(L \to\infty\right) \propto
    \left\{
        \begin{array}{ll}
            L^{-\alpha+1}& \textrm{if $0<\alpha<1$}\\
            const &= \frac{C_V(d)\alpha}{C_A(d)(\alpha - 1)} \qquad \textrm{if $1 < \alpha $}
        \end{array}
    \right.
    \label{eq:divergences}
\end{equation}
as $r_{max}$ si related to the system size $L$, by $L = r_{max}\cdot R_S$.
Formulas (\ref{eq:eps}) and  (\ref{eq:divergences}) evidence that for 
$\alpha<1$ if the average distance between the disks is kept constant, then 
the turbid fraction is vanishing with the system size.

\begin{figure}
    \centering
    \includegraphics[width=.95\columnwidth]{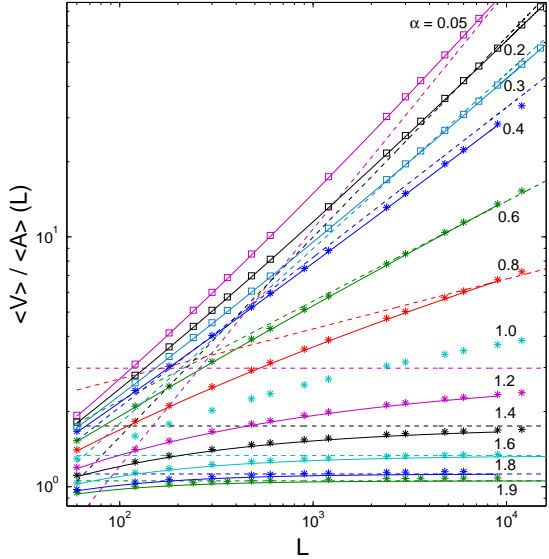}
    \caption{(Color online) The comparison of $V/A(L)$: the experimental results
        (squares and asterisk), the analytic function (solid line) and the
        $L\to\infty$ limit (dashed line) as found for $0.05\le\alpha\le1.9$.
        The discrete sampling reproduces correctly the analytical result. We also
        note that, for  $0.05\lesssim\alpha\lesssim 0.4$ and for
        $0.6\lesssim\alpha\lesssim 1.8$ we are in a pre-asymptotic range, and the $V/A$
        ratio has not converged yet to the expected value. In particular, for low $\alpha$
        finite size effects overestimate the value of the exponent while at
        large $\alpha$ the value is underestimated.
    }
    \label{fig:V_A_asymptotic}
\end{figure}

\begin{figure}
    \centering
    \includegraphics[width=.9\columnwidth]{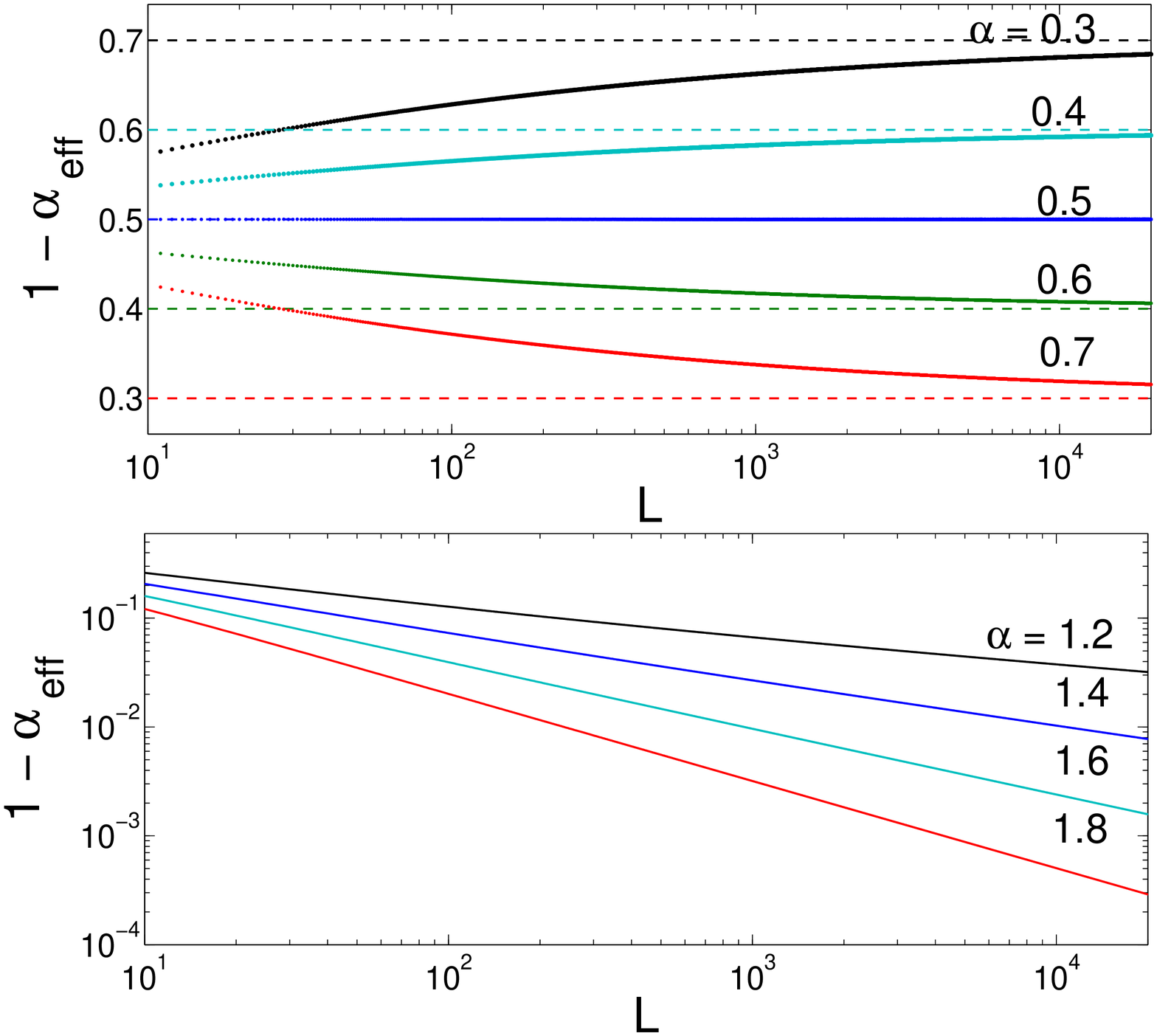}
    \caption{
        (Color online) (top) The $1 - \alpha_{eff}(L)$ curves for $0<\alpha<1$. Note that for
        $\alpha<1/2$ we find $\alpha_{eff}>\alpha$ while $\alpha_{eff}<\alpha$
        for $\alpha>1/2$. For $\alpha=0.5$ we have $\alpha_{eff}=\alpha$ for
        every $L$.
        In all the other case the $1 - \alpha_{eff}(L)$ functions slowly converge to the
        asymptotic limit $1 - \alpha$ (dashed lines).
        (bottom) The $1 - \alpha_{eff}(L)$ curves for $1<\alpha<2$. Again, we found 
        slow convergence to the asymptotic limit $1 - \alpha_{eff} = 0$.
    }
    \label{tab:alpha_eff}
\end{figure}

We now compare in Fig. \ref{fig:V_A_asymptotic} $V/A(L)$ as found by using our
simulations, and  the analytical results of Eq. (\ref{eq:analyt_finite}),
showing also the thermodynamic limit $L\to\infty$ as found in Eq.
(\ref{eq:divergences}). Fig. \ref{fig:V_A_asymptotic}  evidences that the
sampling correctly reproduces the analytical form of $V/A(L)$.
However, for  $0.05\lesssim\alpha\lesssim 0.4$ and for $0.6\lesssim\alpha$, this
ratio grows differently with respect to the expected asymptotic behaviour.
This means that the finite system displays an effective exponent 
$\alpha_{eff} \ne \alpha$.
The value of  $\alpha_{eff}$ as a function of the scale $L$ can be easily
calculated by the logarithmic derivative of Eq. (\ref{eq:analyt_finite}):
\begin{equation}
1- \alpha_{eff}=   \frac{d}{d \ln L } \ln \frac{V}{A} (L)  =(1-\alpha) \frac{1}{1-L^{\alpha - 1}} + \alpha \frac{L^{-\alpha}}{L^{-\alpha} - 1}.
    \label{eq:VA_dlog}
\end{equation}
The function is plotted in Fig. \ref{tab:alpha_eff}, showing that for
$\alpha<0.5$ finite size effects overestimate the value of the exponent
($\alpha_{eff}> \alpha$), while for $\alpha>0.5$ we have $\alpha_{eff} <
\alpha$.

Let us now evaluate $\epsilon(L)$ as defined in Eq.  (\ref{eq:eps}), estimating
the average surface, the volume of disks and the filling fraction $f$ in our
packing simulations. In Fig. \ref{fig:epsilon_L} we plot $\epsilon(L)$  for  the
$0.05\le\alpha\le1.8$ range. $\epsilon(L)$ diverges for $\alpha\lesssim 0.5$.
This implies that the packing is not working, since at large values of $L$ one
is not able to keep a constant distance between the disks. Therefore, the
average turbid region between particles has a linear size greater than the
scattering length $\lambda$, so that the light ray will pass more time in the
scattering region, slowing down its dynamic.
On the other hand, $\epsilon(L)$ seems to be constant ($\epsilon\sim 0.4$) for
$\alpha\gtrsim 0.5$. This suggests a real packing with limited space between
disks. In the intermediate regime $\alpha\simeq 0.4-0.5$, $\epsilon(L)$ display
large oscillations evidencing an instability in the packing procedures. We
notice that this crossover region corresponds to the value $\alpha=0.5$ where
$\alpha_{eff}$ as a function of $L$ change its behavior (see figure
\ref{tab:alpha_eff}).

\begin{figure}
    \centering
    \includegraphics[width=.95\columnwidth]{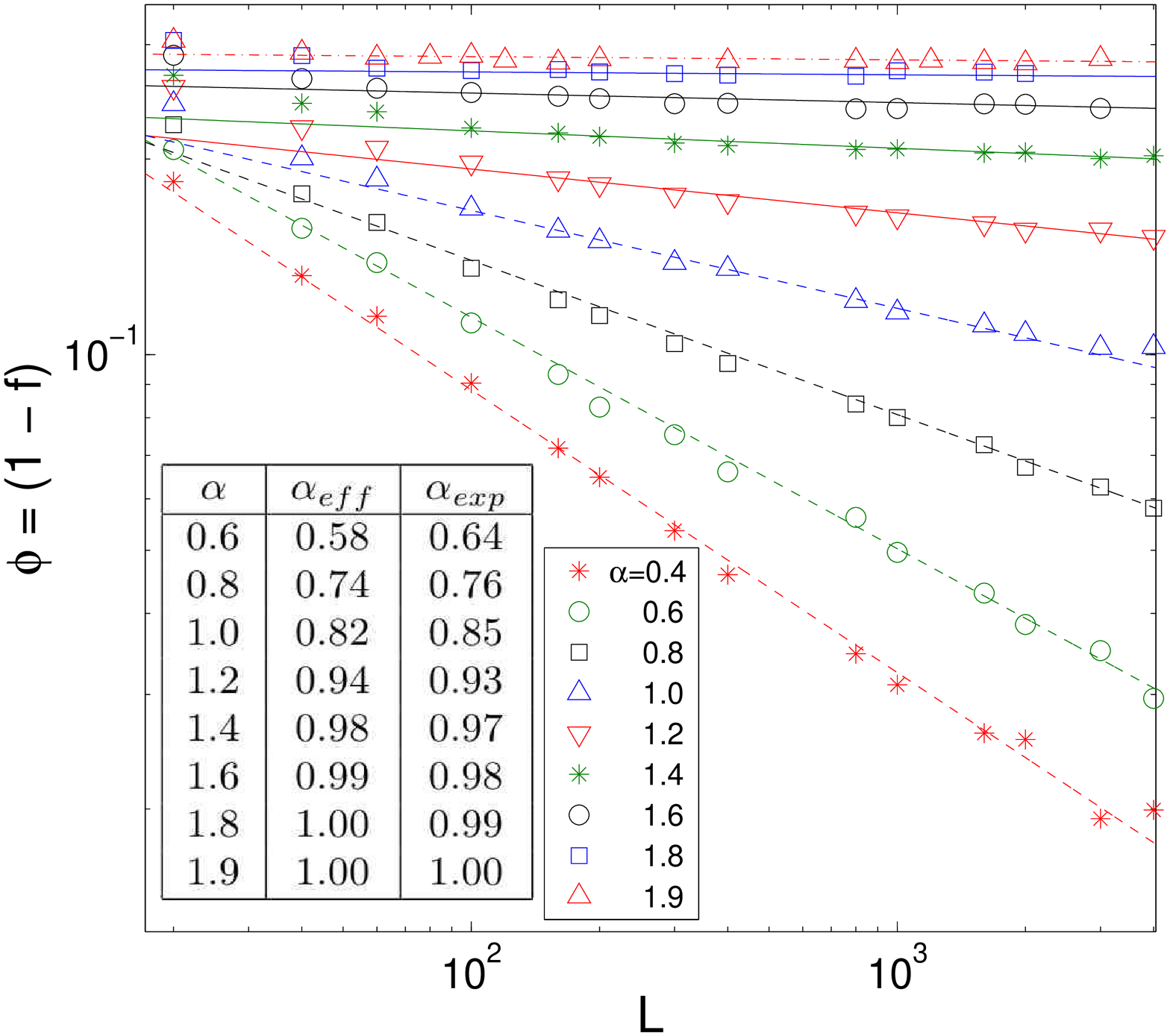}
    \caption{
        (Color online) The fitted $\phi(L)$ curves giving the $\alpha_{exp}$
        exponents.
        \newline
        (Inset) The comparison between the two exponents $\alpha_{eff}$ and
        $\alpha_{exp}$ as found for different $\alpha$ in Fig.
        \ref{fig:alpha_eff_exp}.
    }
    \label{fig:alpha_eff_exp}
\end{figure}

From this plot, it is clear that the region $\alpha\lesssim 0.5$, featuring an
anomalous dynamical behavior in \cite{beenakker:2012}, does not correspond to a
real packing and should be excluded from our measures. On the other hand, once
$\alpha$ exceeds the $\sim 0.4-0.5$ threshold, $\epsilon(L)$ appears to be
constant, so the $(1-f)/f$ term in Eq. (\ref{eq:eps}) has to decrease as
$A/V(L)$. It is then reasonable to compare $\alpha_{exp}$, coming from the
direct measure of the turbid fraction from packing simulations, with
$\alpha_{eff}$ obtained in theoretical analysis.
In the regime where $\epsilon(L)$ is almost constant we should find
$\alpha_{exp}\sim\alpha_{eff}$.
In Fig.\ref{fig:alpha_eff_exp} we present the results of this analysis
evidencing a good agreement.

The general picture seems now quite clear. Although we choose the disks radii 
according to a certain $\alpha$ (or a $\beta = \alpha + d - 1$), at finite
size the packing features a different self similar structure: there
exists another exponent $\alpha_{eff}$ which effectively drives the geometrical
scaling since the thermodynamic limit has not been reached yet.  Moreover, the
mean thickness $\epsilon$ of the turbid region around each sphere has two 
different behavior for $0.1\le\alpha<0.5$ and $0.5\le\alpha<2$, respectively.
In the first range, $\epsilon$ diverges giving rise to an anomaly in the system
topology, and signaling that the packing fails.
On the other hand, when $0.5\le\alpha<2$, $\epsilon$ is almost constant. This in
turn results in the comparable values of the two exponents $\alpha_{exp}$ and
$\alpha_{eff}$, driving the scaling of $\phi\propto L^{\alpha_{exp}-1}$ and
$\frac{V}{A}(L)\propto L^{1-\alpha_{eff}}$. The exponent $\alpha_{eff}$ is
therefore the true geometrical parameter related to the effective packing.

\section{Transport in $2d$ random  L\'evy packings}
\label{sec:dynamics}

\subsection{The total transmission}

\begin{figure}
    \centering
    \includegraphics[width=.95\columnwidth]{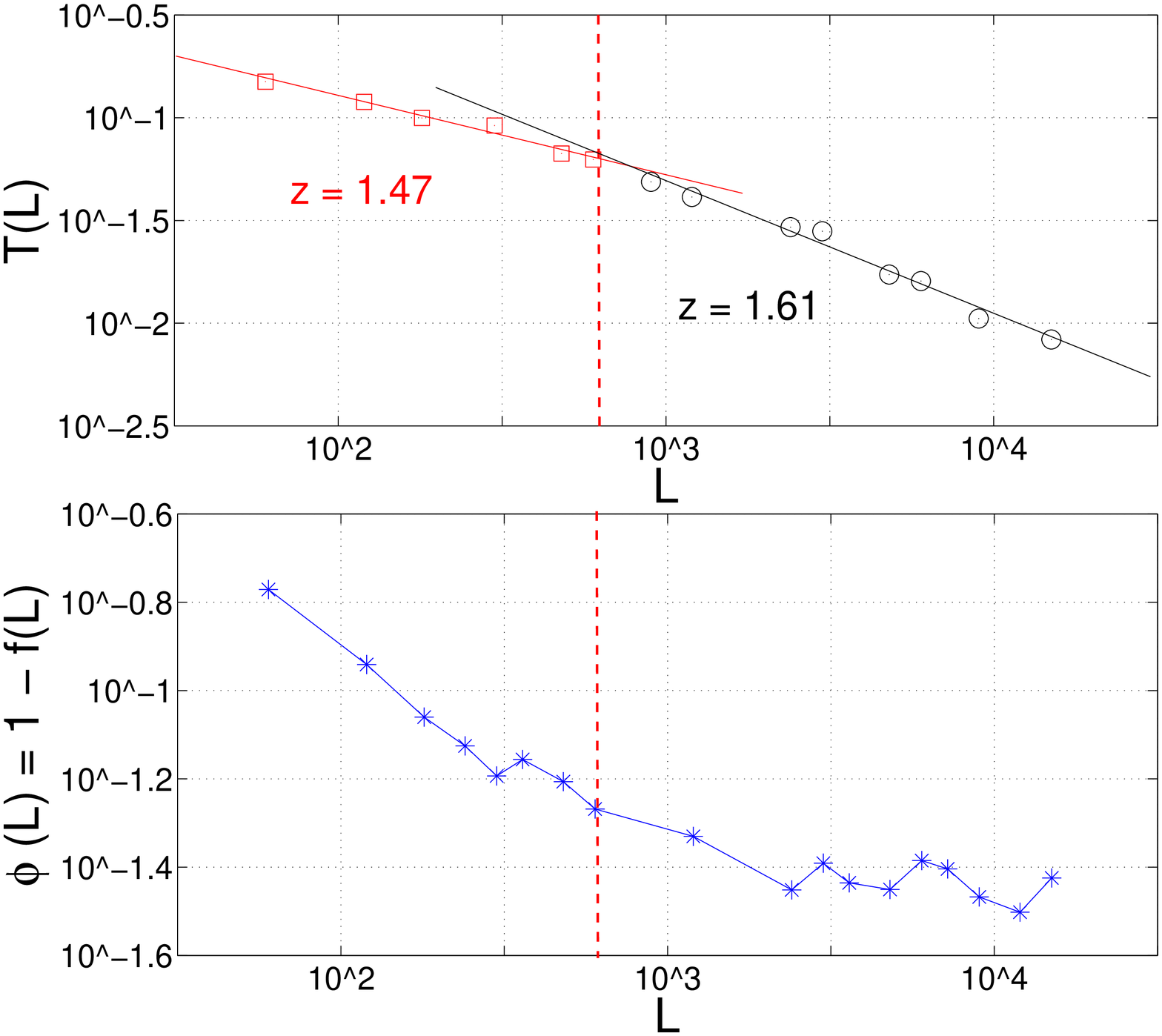}
    \caption{
        (Color online) $\alpha = 0.2$: (Top)
        the $T(L)$ function with the two different interpolating lines (squares
        and circles for the experimental points and the fitting solid lines):
        squares for the low thicknesses range and circles for the higher ones.
        (Bottom) The $1 - f(L)$ (turbid volume) function (asterisks). The
        delimiting thickness $L_D \sim 650$ is enlightened with the vertical dashed line.
    }
    \label{fig:t_tot_vs_ff_a02}
\end{figure}

The first quantity we measure in simulations at optimal filling is the 
transmitted intensity for different system thicknesses.
A ray of light starts from the $x = 0$ surface 
and diffuses according our dynamics 
until it comes back at the $x = 0$ 
(reflection), or it reaches $x=L_x$ (transmission).
We then count the number $r_t(L_x)$ of rays that cross the slab 
on the total number of rays  $r(L_x)$ and we obtain the
transmitted intensity $T(L_x) = r_t(L_x)/r(L_x)$.
Repeating the simulation and varying the thickness of the slab, we eventually find
the $T(L_x)$ function.  We expect to find a scaling of the transmitted
probability $T(L_x)$ function, with now $L_x=L$ following the scaling relation
\cite{Burioni:2010fk,Buonsante:2011}:
\begin{equation*}
    T(L) \propto \frac{1}{L^{(z - 1)}}.
    \label{eq:T_t_sim}
\end{equation*}
In Fig. \ref{fig:t_tot_vs_ff_a02} we plot the resulting $T(L)$
for $\alpha = 0.2$, in the region where packing fails and the distance 
between disks increases with the size.
Clearly, there are two different ranges of the size where both 
the turbid
fraction and the transmission probability behave differently.
For low thickness range ($L \in [60, 650]$) the algorithm succeeds in
filling the system and, in
fact, the turbid fraction decreases as a power law with $L$. Inside this
interval, the packing works and the transmission probability scales with a
super-diffusive exponent $z = 1.47$ (much lower than the $z = 2$ expected for a
diffusive regime). Then, for  $L > 650$, the turbid fraction stops 
lowering with increasing size and  the system presents
an almost constant turbid fraction, i.e. $f \sim 0.96$.
In this regime, the transmission probability behaves differently as well.
We find indeed that the scaling exponent governing the $T(L)$
becomes closer to the diffusive case $z=2$.
The same qualitative behavior characterizes the whole range 
$\alpha \in [0.0 ,0.4]$.

\begin{figure}
    \centering
    \includegraphics[width=.95\columnwidth]{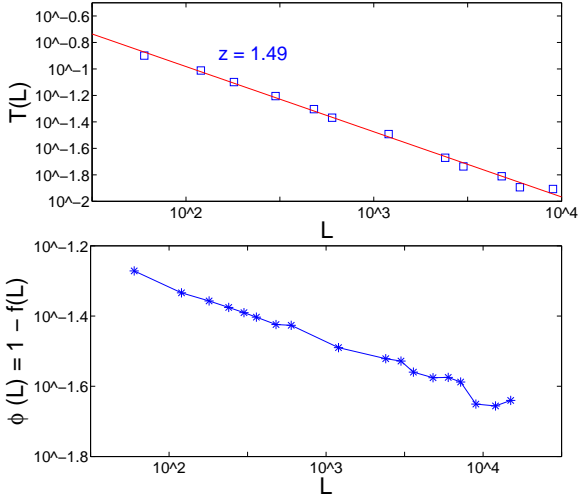}
    \caption{
        (Color online) $\alpha = 0.6$: (Top) The $T(L)$ function (squares) with the scaling
        interpolation (solid line). (Bottom) The $1 - f(L)$ (turbid volume) function
        (asterisks).
    }
    \label{fig:t_tot_vs_ff_a06}
\end{figure}

\begin{figure}
    \centering
    \includegraphics[width=.9\columnwidth]{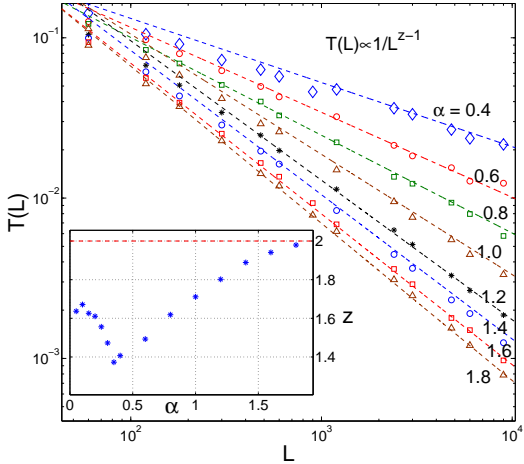}
    \caption{
        (Color online) The transmitted intensity $T(L)$ scaling with system size
        $L$ for different $\alpha$ value. The fit of the function gives the
        exponent $z-1$.
        (Inset) The dynamic exponent $z$ (asterisks) as a function of
        $\alpha$.
        Here, we merged the $z$ found in the anomalous range $0.1\le\alpha\le0.4$.
        The diffusive limit $z=2$ (dashed line) is shown.
    }
    \label{fig:T_L_scale}
\end{figure}

As we can see in Fig. \ref{fig:t_tot_vs_ff_a06}, 
outside this problematic range we
eventually find a power law decrease of the turbid fraction in the whole
range of $L$ we analyzed.
In Fig. \ref{fig:T_L_scale} we show the transmitted intensity for 
$0.4\le\alpha<2$. The fitting curves gives the values for $z$ that are resumed
in the inset of Fig. \ref{fig:T_L_scale}, in the
anomalous range the value of the exponent increases, due to the failure of the 
packing procedure. The plot agrees very well
with that presented for the $2d$ case in \cite{beenakker:2012}.

\subsection{The Time resolved transmission}
\label{ssub:Time Resolved}
\begin{figure}[h!]
    \centering
    \includegraphics[width=.95\columnwidth]{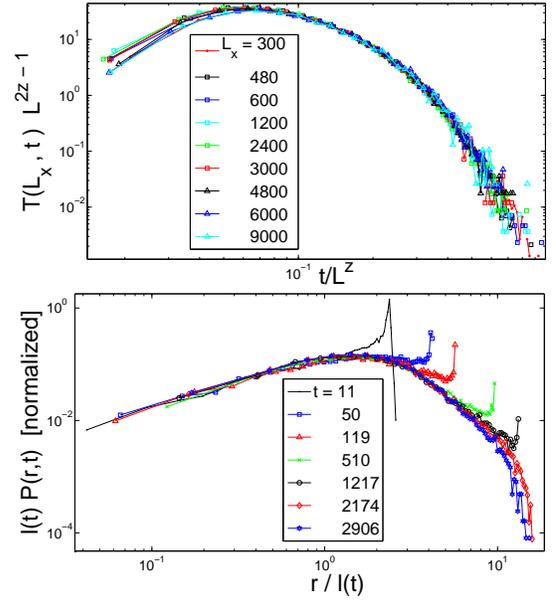}
    \caption{
        (Color online) (a) Scaling of time-resolved transmission $T(L,t)$ for $\alpha = 1.2$
        and different slab thicknesses.
        The characteristic length is set to $L^{z}$, as in Eq.
        (\ref{eq:scaling_Tlt}).
        (b) Scaling of the average probability $P(r,t)$ for the system with
        $\alpha=0.8$ and $L = {r_max} \cdot R_S = 6\cdot10^3$. The spikes stem
        from the ballistic motion inside the bigger disks and they disappear as
        soon as $t\gtrsim r_{max}$.
        The scaling pictures holds in the whole time range analyzed, satisfying
        the scaling hypothesis made in Eq. (\ref{eq:prt_scaling}). The
        characteristic length has been set to $l(t)=t^{1/z}$, being $z$ the
        dynamical exponent governing the transmitted intensity.
    }
    \label{fig:prt_T_t}
\end{figure}

The next dynamical quantity we analyze is the time-resolved transmission.
The parameters of the simulations are identical to the ones outlined before.
The transmission (or backscatter) time of each particle is recorder and binned
in a histogram, and the time resolved transmission should follow the scaling
form:
\begin{equation}
    T(t,L) = L^\eta  \tilde f(L/l(t)),
    \label{eq:scaling_Tlt}
\end{equation}
while the $P(r,t)$ the  probability function, should scale as
\begin{equation}
    P(r,t) = l^{-1}(t)f(r/l(t)).
    \label{eq:prt_scaling}
\end{equation}
We are now able to evaluate both scalings, as the growth of the characteristic
length is related to the total transmission by the Einstein relation
\cite{Burioni:2010fk,cates}
$l(t)\propto t^{1/z}$ and $\eta=1-2z$. As we show in Fig. \ref{fig:prt_T_t}, the scaling picture
holds for both for $P(r,t)$ and $T(L,t)$, evidencing a nice data collapse at
least at large enough $L$.

\subsection{Scaling and the effective $\alpha$ exponent}
\label{ssub:alpha_eff}

Our simulations confirm the results in \cite{beenakker:2012} that 
superdiffusive anomalous transmission is observed for
$0.5\lesssim \alpha \lesssim 1.6$, at variance with the deterministic
self-similar models of \levy packings, where super-diffusive behavior 
occurs only for $\alpha<1$ \cite{Buonsante:2011}.
Interestingly, in the same regime $0.5 \lesssim \alpha \lesssim 1.6$ 
our static study evidences
that, for finite size systems, the filling fraction
is not described by the exponent $\alpha$ characterizing the radii distribution
$p(r)\sim 1/r^{d+\alpha}$ but an effective size dependent exponent
$\alpha_{eff}$ has to be introduced.
We, therefore, expect that also the transport properties may be affected by 
the finite size of the system.

\begin{figure}
    \centering
    \includegraphics[width=.9\columnwidth]{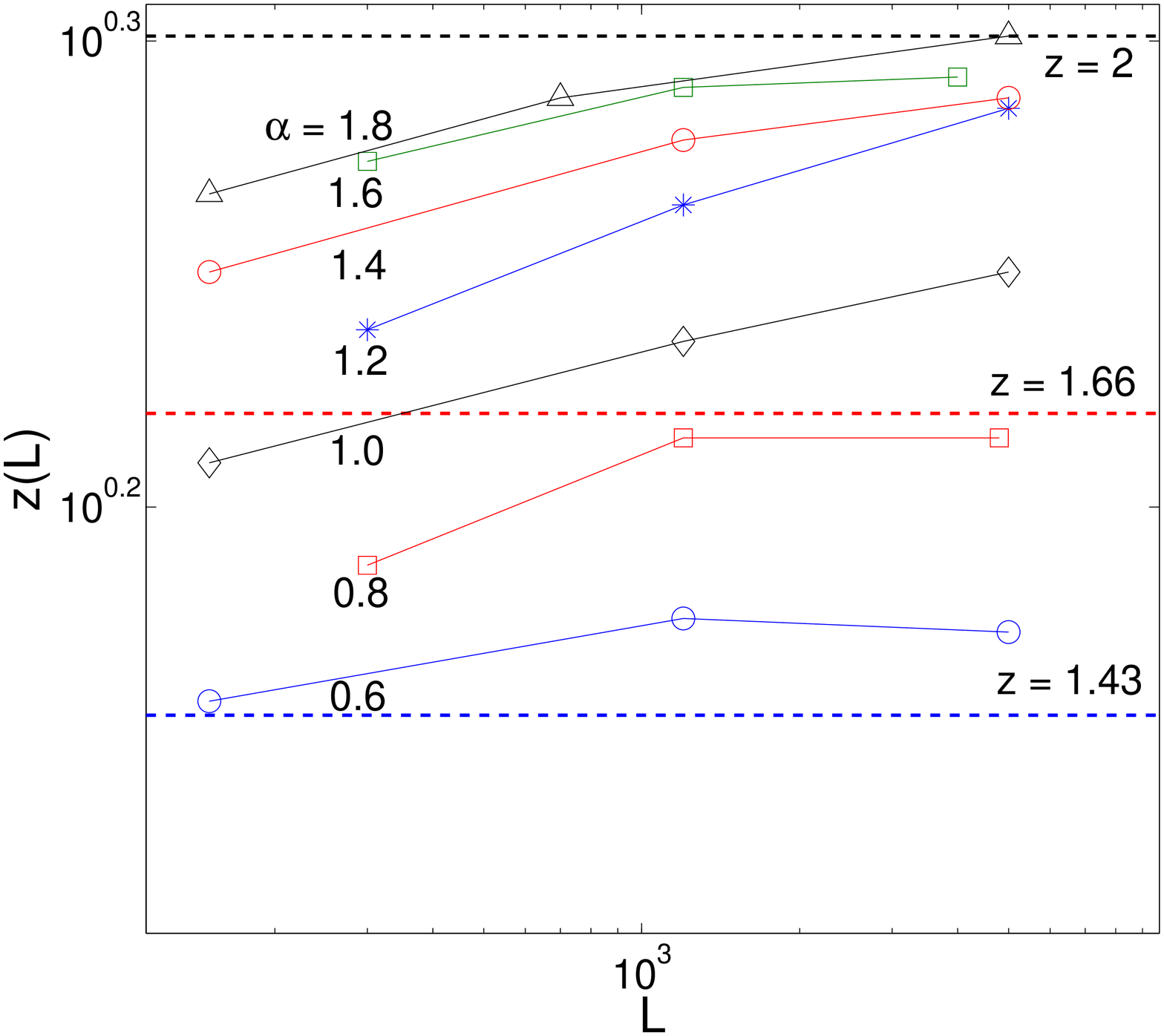}
    \caption{
        (Color online) The $z(L)$ function. The data refer to the  $\lambda=1$ and
        $0.6\le\alpha\le1.8$ simulations. The $z$ suggested by the ansatz
        $z=2/(2-\alpha)$ ($\alpha<1$) and the diffusive limit $z=2$
        ($\alpha\ge1$) are shown for comparison (dashed lines).
        While the exponents found for $\alpha<1$ seem to stop their rise with
        $L$, for $\alpha\ge1$ they slowly grow toward $z=2$. The latter is
        reached only by the $\alpha=1.8$ set (black triangles) at the largest
        system size analyzed.
    }
    \label{fig:z_L_fmax}
\end{figure}

We now calculate the exponent $z$ for different values of $L$. In particular, we
estimate $z(L)$ by fitting consecutive intervals of the $T(L)$ function
separately. For example, the fitting of $T(L)$ in the $60\le L\le 400$ range
provides the value of $z$ at the average length of the analyzed range $L=230$.
In Fig. \ref{fig:z_L_fmax} we show the results of this analysis.

In general $z(L)$ features a growth with the system size and its limit is
consistent with the value $z=2$ for $\alpha>1$, while for $\alpha<1$ also in the
extrapolated infinite size regime a superdiffusive $z$ seems to persists.
This behavior  at large $L$ is similar to the case of the deterministic \levy
fractals. We notice that in the simulation in \cite{Buonsante:2011}, due to the
deterministic rule used to build the fractals, much larger systems can be
considered and  finite size effects are in general negligible.

\begin{figure}
    \centering
    \includegraphics[width=.9\columnwidth]{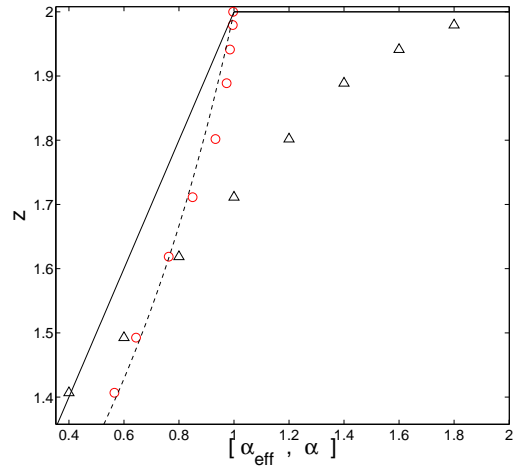}
    \caption{
        (Color online) The comparison between $z(\alpha_{eff})$ (circles), and $z(\alpha)$
        (triangles).
        The $1D$ analytical result $z=\alpha+1$ (black solid line) and the
        ansatz $z=\frac{2}{2-\alpha}$ (black dashed line) are also shown.
        The $z(\alpha)$ significantly differs from the deterministic fractals case
        ansatz and from the $1D$ analytical result. Once we consider $\alpha_{exp}$
        as the system characteristic exponent, the dynamical exponents $z$ appear
        to recover the curve found in the deterministic fractals case.
    }
    \label{fig:z_alpha_eff}
\end{figure}

At finite size, it is therefore reasonable to study the exponent $z$ as a
function of the effective exponent $\alpha_{eff}$ characterizing the finite size
packing, and  compare this function with the one found in \cite{Buonsante:2011}
on a deterministic packing. Let us recall that in that work $z(\alpha)$ follows
the heuristic ansatz $z=2/(2-\alpha)$ in the $0<\alpha\le1$ range, while $z=2$
for $\alpha >1$.
In Fig. \ref{fig:z_alpha_eff} we show that indeed $z(\alpha_{eff})$ is well
approximated by this ansatz ($z=2/(2-\alpha)$ is also plotted in Fig.
\ref{fig:z_L_fmax}). The diffusive regime $z=2$ is reached only when
$\alpha_{eff}\to1$, recovering the distinction between a superdiffusive and
diffusive regime below and above $\alpha=1$ respectively.
Our analysis have been performed in the $2d$ case, but we expect this
phenomenology to hold also in $3d$ systems.

\subsection{The scattering mean free path }
\label{sec:MFP}
The dynamical exponents in the truly asymptotic regime, at infinite size, do not change
by varying the scattering mean free path $\lambda$.
However it is not clear what is the role of $\lambda$ in a regime where 
preasymptotic effects determine the dynamical behavior
at finite sizes.
Here, we will show that the results obtained in previous sections, at least at
large $L$, are robust with respect to a variation of the scattering mean free
path. We remark that we set $\lambda=1=r_{min}$ as in \cite{beenakker:2012}. 
However,  in the experiments
$r_{min}=2.5\mu m$ and $\lambda=12\mu m$, thus corresponding to $\lambda\sim5$
in our setup.

\begin{table}
    \centering
    \begin{tabular}{|c|c|c|c|}
        \hline
        {Parameters}  &   $\lambda=10$ &   $\lambda=1$     &   $\lambda=0.1$ \\
        \hline
        $\alpha=0.8$ &   $1.61$  &   $1.64$  &   $1.64$     \\
        \hline
        $\alpha=1.0$ &   $1.76$  &   $1.78$  &   $1.79$     \\
        \hline
        $\alpha=1.2$ &   $1.89$  &   $1.90$  &   $1.97$     \\
        \hline
    \end{tabular}
    \caption{
        The dynamical exponents $z$ found by varying $\lambda$.
    }
    \label{tab:z_mfp_max}
\end{table}

\begin{figure}
    \centering
    \includegraphics[width=.9\columnwidth]{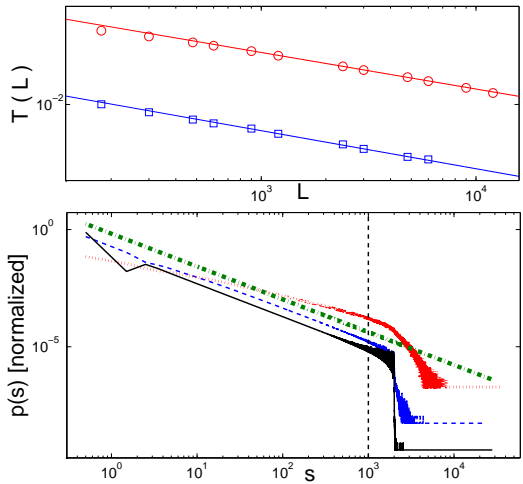}
    \caption{
        (Color online) (top) The transmission probability for $\lambda = 0.1$ (squares) and
        $\lambda = 10$ (circles) for $\alpha=1.0$. The fitting lines are reported,
        giving $z = 1.76$ and $z = 1.79$ for the $\lambda = 10$ and $\lambda =
        0.1$ respectively.
        We fitted taking into account the largest system sizes to avoid,
        especially in the $\lambda=10$ case, the multi-disks crossing events to
        influence the $T(L)$ estimation.
        (bottom) The $p_s(s)$ single step length probability function for $\lambda=0.1$
        (solid line), $1$ (dashed line) and $10$ (dotted line). All the sets are for $\alpha=1.0$,
        $L_x=6\cdot10^3$ and $R_S = 6$. The dot-dashed line shows the expected
        $p_(s)\sim s^{-(\alpha+1)}$ for comparison.
}
    \label{fig:ps_3lambda}
\end{figure}

\begin{figure}
    \centering
    \includegraphics[width=.9\columnwidth]{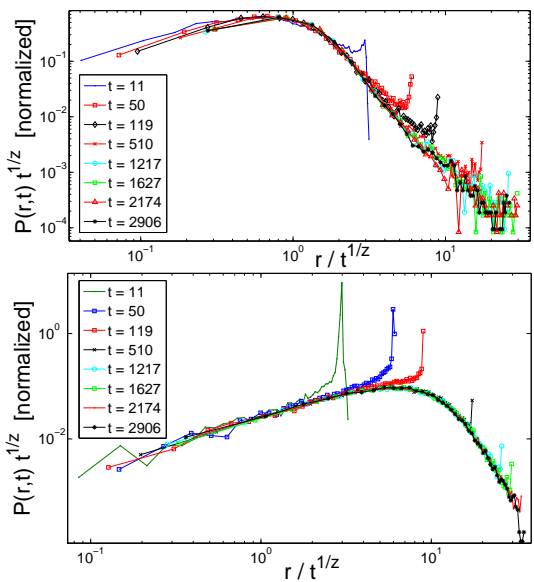}
    \caption{
        (Color online) (top) The scaling of $P(r,t)$ as from the simulation with $\lambda=0.1$,
        $\alpha=1.2$ and $L = r_{max}\cdot R_S = 6\cdot10^3$. The scaling length
        is set to $l(t)=t^{1/z}$ with $z(\alpha=1.2)=1.94$;
        (bottom) The same function has been plotted for the system featuring
        $\lambda=10$. Here $z(\alpha=1.2)=1.93$.
}
    \label{fig:prt_lambda}
\end{figure}
We analyze the system response to a variation of  $\lambda$, keeping $R_S
= 6$,  and $\alpha=[0.8,1.0,1.2]$ and averaging over $5\cdot10^6$ initial
conditions.
Our fits at large $L$ show  that $z$ does not change switching from
$\lambda=0.1$ to $10$ and remains in good agreement with the case $\lambda=1$
(see Table (\ref{tab:z_mfp_max})) though there is, of course, a drop in the
transmission probability $T(L)$ (see Fig. \ref{fig:ps_3lambda} upper panel).
As we can see in Fig. \ref{fig:prt_lambda} the scaling of the $P(r,t)$ 
works in both cases, $\lambda = 0.1$ and $\lambda=10$ once the characteristic
length $l(t)$ is taken into account.

The main differences at  varying $\lambda$ regard the frequency of multi-disks
crossing events and they are evidenced in Fig. \ref{fig:ps_3lambda}, lower panel,
by inspecting the single step length distribution $p_s(s)$. We note indeed that
the bump, found for $\lambda=1$ in the range $1\lesssim s\lesssim 20$, 
disappears for $\lambda=0.1$. This is a clear indication that we are
isolating the chord distribution of the disks, avoiding multi-disks crossing.
We remark that the average distance between the spheres in our packing is 
$\epsilon\simeq 0.4$ (see Figure \ref{fig:epsilon_L}) and  multi-disks crossing 
is completely avoided only for $\lambda \ll \epsilon$.
On the contrary, for $\lambda=10$ the bump spreads over the whole step size
range, even further than $s \sim r_{max}$. Here the multi-disks
crossing events are the leading process within the slab. The signal dies very
slowly and we cannot distinguish a range where $p_s(s)$ follows the chords
distribution function.
This is at variance with the $\lambda=[0.1, 1]$ systems, where a sharp cut-off
is present at the $s\sim 2r_{max}$ length.

\subsection{The truncation length}
\label{sec:exp_geometry}

As a last check of the lengths involved in our model, we consider explicitly the
effects of truncation in the radii distribution i.e. of the parameter
$R_S$.
We run an additional set of simulations for $R_S=[2,\,4,\,6]$,
maximized $f$, $\lambda = 1.0$ and $0.4\le\alpha\le2.8$;

In Table (\ref{tab:alpha_exp_2_4_6}) we show the results for the effective
exponent $\alpha_{eff}$ evidencing that systems with different $R_S$ features
the same $\alpha_{eff}$ i.e. they are equivalent from a geometrical point of
view.
We then compute the transmission properties. The resulting $z$ are shown in
Table (\ref{tab:z_d2r}).

\begin{table}
    \centering
    \begin{tabular}{|c||c|c|c|c|c|c|c|c|}
        \hline
        {{$R_S$}/{$\alpha$}} & $0.4$ & $0.6$ & $0.8$ & $1.0$ & $1.2$ &
        $1.4$ & $1.6$ & $1.8$ \\
        \hline
        \hline
        $2$ & $0.54$ & $0.62$ & $0.74$ & $0.85$ & $0.92$ & $0.95$ & $0.98$ &
        $0.99$ \\
        $4$ & $0.53$ & $0.66$ & $0.78$ & $0.86$ & $0.96$ & $0.98$ & $0.99$ &
        $1.00$ \\
        $6$ & $0.56$ & $0.64$ & $0.76$ & $0.85$ & $0.93$ & $0.97$ & $0.98$ &
        $1.00$ \\
        \hline
    \end{tabular}
    \caption{
        The $\alpha_{eff}$ exponents found for different $R_S$.
    }
    \label{tab:alpha_exp_2_4_6}
\end{table}

\begin{table}
    \centering
    \begin{tabular}{|c||c|c|c|}
        \hline
        $\alpha$   &   $z$ ($R_S=2$)    &    $z$ ($R_S=4$)  &   $z$ ($R_S=6$) \\
        \hline
        \hline
        $0.4$   &   $1.43$  &   $1.36$  &   $1.41$  \\
        $0.6$   &   $1.49$  &   $1.52$  &   $1.55$  \\
        $0.8$   &   $1.55$  &   $1.65$  &   $1.65$  \\
        $1.0$   &   $1.67$  &   $1.77$  &   $1.76$  \\
        $1.2$   &   $1.72$  &   $1.90$  &   $1.91$  \\
        $1.4$   &   $1.70$  &   $1.94$  &   $1.93$  \\
        $1.6$   &   $1.72$  &   $1.96$  &   $1.96$  \\
        $1.8$   &   $1.75$  &   $1.98$  &   $2.00$  \\
        \hline
    \end{tabular}
    \caption{
        The resulting dynamical exponents $z$ for the $L=2r_{max}$
        ($R_S=2$) and $L = 4r_{max}$ ($R_S=4$) cases compared with the
        $R_S = L/r_{max}=6$ case.
    }
    \label{tab:z_d2r}
\end{table}

\begin{figure}
    \centering
        {\includegraphics[width=.9\columnwidth]{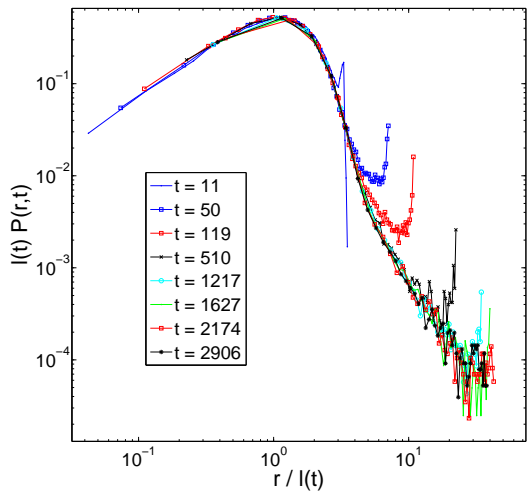}}
    \caption{
        (Color online) Scaling of the $P(r,t)$ function for a $R_S=2$ sample with      $L=2000$
        and $\alpha=1.0$. The characteristic length is $l(t)=l^{1/z}$.
        The spikes correspond to the ballistic motion within the largest disk in
        the slab. The latter is now as large as the slab itself.
           }
    \label{fig:res_d2r_fig}
\end{figure}

The dynamical exponents $z$ seem to be independent of the truncation for
$R_S>2$, while a discrepancy is found for what concerns the data at $R_S = 2$
especially in the regime of large $\alpha$ i.e. $\alpha>0.8$.  In particular,
the dynamical exponent $z$ referring to $R_S = 2$ is found to be smaller than
the one computed for $R_S = 6$ and $R_S = 4$. We remark that for $R_S=2$ the
diameter of the largest sphere equals the system size $L$, and direct ballistic
transmission becomes much more likely.
Nevertheless, Fig. \ref{fig:res_d2r_fig} evidences that even for $R_S=2$ the
scaling scheme for the probability distribution $P(r,t)$  holds using the proper
characteristic length $l(t)=t^{1/z}$.

\section{\levy Packings at fixed filling}
\label{sec:experiment}

The experimental setup described in \cite{Barthelemy:2008rt}
presents  an important difference with respect to our simulations
\ref{sec:dynamics}. The theoretical packings, in order to reproduce  a fractal
sampling, try to maximize the filling fraction $f$, i.e. the number of disks in
the slab. On the opposite, in the experiments the filling fraction is kept
constant, in particular $f=71\%$. Therefore, in the experiment, $\alpha$ only
affects the step length distribution, but plays no role in the behavior of $f$
passing from one system scale to another.
Recalling the packing procedure, we notice that if the term $(1-f)/f$ in Eq.
(\ref{eq:eps}) is constant with respect to $L$, then at $\alpha<1$ the packing
features a diverging sphere-to-sphere distance $\epsilon(L)$. We then expect to
observe a diffusive behavior as the size $L$ grows, since the diffusive region
becomes dominant. However, at finite size the average spacing $\epsilon$ between
the spheres can be comparable to $\lambda$, and this could give rise also in
this case to an effective exponent $z$ smaller than 2, since at that scale the
underlying disks distribution display a fractal superdiffusive geometry.

For this purpose we now analyze finite size effects in a system with fixed
$f=0.7$, varying $\lambda=[0.1,1.0,10]$  and $\alpha=[0.8,1.0,1.2]$.
To construct a slab with fixed  $f$, we increment the number of disks until we
reach the desired value of $f$. The disk are then placed randomly according the
usual algorithm. We found that $f\sim0.7$ is a filling fraction value that can
be easily reached at all the system sizes and for all the $\alpha$ exponents, so
that the samples are created very quickly.
We simulated the walk of $5\cdot10^6$ particles, analyzing the transmission
properties of the system. The geometry is still fixed to $R_S=6$ and
$r_{min}=1$.

\begin{figure}
    \centering
    \includegraphics[width=.9\columnwidth]{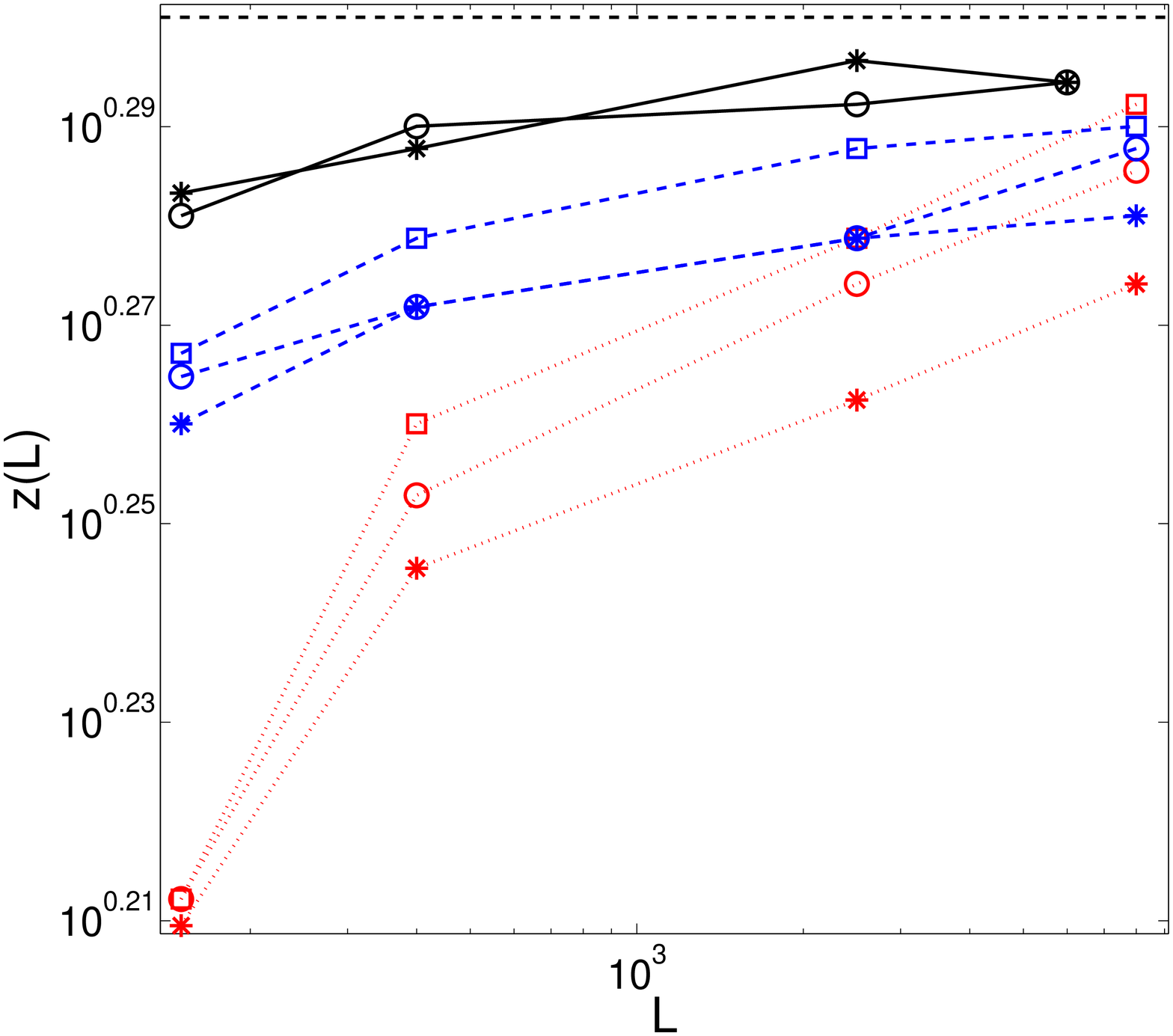}
    \caption{
        (Color online) The evaluation of $z(L)$ for system with $f_{fix}=0.7$. The data
        refer to $\lambda=10$ (dotted lines), $\lambda=1$ (dashed lines) and $\lambda=0.1$
        (solid lines). The system exponents are $\alpha=0.8$ (asterisks),
        $\alpha=1.0$ (circles) and $\alpha=1.2$ (squares) respectively.
        The $z(L)$ found for $\lambda=[1,10]$ seems to converge to the
        exponents found in the $\lambda=0.1$ case, which is almost constant
        over the whole range.
        The black dashed line shows the diffusive limit $z=2$.
    }
    \label{fig:z_L_fix}
\end{figure}

We test the scaling with the system size by measuring $z(L)$ obtained by fitting
$T(L)$ on different $L$-intervals.
The results are plotted in Fig. \ref{fig:z_L_fix}.
The $z$ exponent is converging to the diffusive case as the system size increases.
Furthermore, the data from  $\lambda=1$ and $\lambda=10$ undergo a rapid growth,
whereas the $\lambda=0.1$ exponent is slowly increasing, as if it has already
reached the diffusive case.
However, all the exponents seem to converge to a common $z$ value, very close to
the diffusive $z=2$.

The results obtained at the maximum size $L = 1.2\cdot10^4$ 
are summarized in Table \ref{tab:z_tutti_casi}.
The data confirm that the convergence to $z=2$ is faster for smaller values of
$\lambda$. For instance, in the case $\alpha=0.8$ we find $z=1.86$ and $z=1.97$
for $\lambda$ set to $10$ and $0.1$, respectively. Moreover, if we lower the 
filling fraction $f_{fix}$ to $0.5$, $\epsilon(L)$ is, obviously, growing faster with 
the system size and the dynamics converges more rapidly to a diffusive regime.

\begin{table}
    \centering
    \begin{tabular}{|c|c|c|c|c|}
        \hline
        \multicolumn{2}{|c|}{Parameters} &     $f_{fix}=0.7$     &   $f_{fix}=0.5$ \\
        \hline
        $\lambda=10$    &   $\alpha=0.8$ &   $1.86$  &   $-$     \\
        $''$            &   $\alpha=1.0$ &   $1.94$  &   $-$     \\
        $''$            &   $\alpha=1.2$ &   $1.97$  &   $-$     \\
        \hline
        $\lambda=1$     &   $\alpha=0.8$ &   $1.92$  &   $1.97$  \\
        $''$            &   $\alpha=1.0$ &   $1.95$  &   $1.97$  \\
        $''$            &   $\alpha=1.2$ &   $1.94$  &   $1.96$  \\
        \hline
        $\lambda=0.1$   &   $\alpha=0.8$ &   $1.97$  &   $-$     \\
        $''$            &   $\alpha=1.0$ &   $1.96$  &   $-$     \\
        $''$            &   $\alpha=1.2$ &   $2.00$  &   $-$     \\
        \hline
    \end{tabular}
    \caption{
        The dynamical exponents found by maximizing $f$ ($f_{max}$ column) and by fixing it ($f_{fix}$) columns.
    }
    \label{tab:z_tutti_casi}
\end{table}

\begin{figure}
    \centering
    \includegraphics[width=.8\columnwidth]{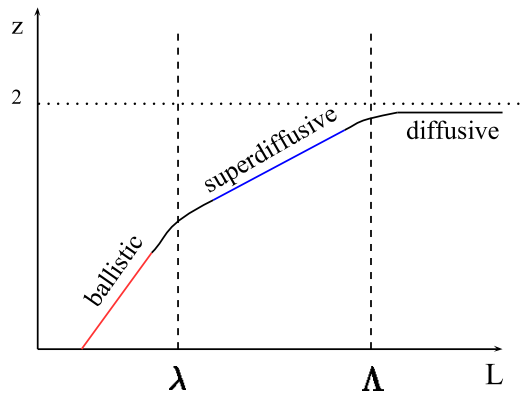}
    \caption{(Color online) The hypothetical convergence of the dynamical exponent $z$ to a diffusive case as we increase the system size.}
    \label{fig:crossing_lengths}
\end{figure}

It is then very likely that we are observing a crossing between a superdiffusive
and diffusive regime. The latter could be driven by a characteristic length of
the system that, so far, is unknown.
This length could depend on the scattering mean free path, the truncation
length, the average disks-to-disk distance (our $\epsilon$) and on the system
exponent $\alpha$ as well, so that the most general form of this length should
be $\Lambda(\alpha,f,\epsilon)$.
As we demonstrated, it is reasonable to expect that decreasing the $\lambda$
length and-or decreasing the $f_{fix}$ value, one should lower this length, so 
to observe at lower sizes the crossing from a superdiffusive to a
diffusive regime. The behaviour of the system is outlined in Fig.
\ref{fig:crossing_lengths}.
We start with a \levy packing whose thickness is equal (or of the same order) of
the scattering mean free path.
Obviously, transport properties follow a ballistic behavior, as the whole slab
is crossed with few jumps. As we increase the system size, we turn the ballistic
behavior in a super-diffusive one until the slab thickness is of the same order
the characteristic length $\Lambda(\alpha,f,\epsilon)$. We are probably
evaluating our dynamic exponents in this central size range, as they converge to
the diffusive case $z=2$ when decreasing $\lambda$ (i.e. when we decrease
$\Lambda$, thus anticipating the crossing). From there on, the system undergoes
an additional crossing to a diffusive case, where the correct, diffusive
exponent is recovered.

As a last step for the fixed filling case, we consider explicitly the effect of
the building procedure of the actual samples, that limits $L$ to be the double
of $r_{max}$, i.e.$R_S=2$.
We run an additional set of simulations: $R_S=[2,\,4,\,6] = L/r_{max}$, fixed $f
= 0.7$, $\lambda = 1.0$ and $0.4\le\alpha\le2.8$; and we compute the
transmission properties. We recover the same behaviour as that found in the
case of optimized filling. In the Table  \ref{tab:z_d2r_f_fix} we resume the
resulting dynamical exponents.
Again, we find good agreement between the $R_S = [4,\,6]$ configurations, while
the $R_S = 2$ systems are not converging to the diffusive limit $z = 2$
since the truncation length is as large as the system size.

\begin{table}
    \centering
    \begin{tabular}{|c||c|c|c|}
        \hline
        $\alpha$   &   $z$ ($R_S=2$)   &    $z$ ($R_S=4$)  &   $z$ ($R_S=6$) \\
        \hline
        \hline
        $0.4$   &   $1.89$  &   $1.94$  &   $-$     \\
        $0.6$   &   $1.82$  &   $1.91$  &   $-$     \\
        $0.8$   &   $1.82$  &   $1.98$  &   $1.92$  \\
        $1.0$   &   $1.84$  &   $1.92$  &   $1.95$  \\
        $1.2$   &   $1.80$  &   $1.98$  &   $1.94$  \\
        $1.4$   &   $1.75$  &   $-$     &   $-$     \\
        $1.6$   &   $1.80$  &   $-$     &   $-$     \\
        \hline
    \end{tabular}
    \caption{
        The resulting dynamical exponents $z$ for the $L=2r_{max}$
        ($R_S=2$) and $L = 4r_{max}$ ($R_S=4$) cases compared with the
        $R_S = L/r_{max}=6$ case. The data refer to the $f_{fix} = 0.7$ case.
    }
    \label{tab:z_d2r_f_fix}
\end{table}

\section{Conclusions}
\label{sec:conc}
Building a tunable medium with given superdiffusive properties is an extremely
interesting task. It can help to unravel the effect of quenched disorder in
presence of large fluctuations and it gives access to the engineering of disordered
media with desired transport effects\cite{Barthelemy:2008rt}. Besides, it can also allow to
extract valuable information on transport and diffusion in natural porous media
\cite{PhysRevLett.65.2201,Palombo:2011mz}. A self similar geometry with a \levy like step length distribution
in a wide range appears to be the crucial ingredient to obtain the desired
superdiffusive effects. A packing obtained at fixed filling fraction can present
a self similar region in a restricted sizes window but at larger sizes it will
turn towards a diffusive sample.

In this paper, we have analyzed in details a $2-$dimensional packing of disks
with \levy distributed radii and we studied the finite size effects arising from
the complex polydispersed disks packing. We have evidenced that the behavior of
the filling fraction at varying system size can be used as a key parameter for
the scaling property of the total transmission and for the time resolved
transmission. The packing at finite size features an effective step length
distribution whose parameter $\alpha_{eff}$ is different from the initial
$\alpha$ extracted from the disks radii self similar distribution.
Superdiffusive effects are then observed when  $\alpha_{eff} <1$. Interestingly,
the exponent $z$  of the scaling length $l(t)=t^{1/z}$ as a function of
$\alpha_{eff}$ is consistent with the ansatz found in deterministic packings, a
result that certainly deserves further investigations.

\begin{acknowledgments}
We wish to acknowledge Romolo Savo, Tomas Svensson, 
Kevin Vynck and Diederik S. Wiersma for fruitful
discussions.
\end{acknowledgments}


\begin{thebibliography}{}
    \bibitem{PhysRevLett.65.2201} A.~Ott, J.~P. Bouchaud, D.~Langevin, and
        W.~Urbach, \newblock {\em Phys. Rev. Lett.} \textbf{65}, 2201 (1990).

    \bibitem{Davis:2002ys} A.~B. Davis and A.~Marshak, \newblock {\em Journal of
        the Atmospheric Sciences} \textbf{59}, 2713 (2002).

    \bibitem{Benson:2001zr} Benson, David A. and Schumer, Rina and Meerschaert,
        Mark M. and Wheatcraft, Stephen W., \newblock {\em Transport in Porous
        Media} \textbf{42}, 211 (2001).

    \bibitem{Palombo:2011mz} M.~{Palombo}, A.~{Gabrielli}, S.~{de Santis},
        C.~{Cametti}, G.~{Ruocco}, and S.~{Capuani}, \newblock {\em J. Chem.
        Phys.} \textbf{135}, 034504 (2011).

    \bibitem{Brockmann:2003uq} D.~{Brockmann} and T.~{Geisel}, \newblock {\em
        Phys. Rev. Lett.} \textbf{90}, 170601 (2003).

    \bibitem{PhysRevLett.72.203} F.~Bardou, J.~P. Bouchaud, O.~Emile, A.~Aspect,
        and C.~Cohen-Tannoudji, \newblock {\em Phys. Rev. Lett.} \textbf{72}, 203
        (1994).

    \bibitem{PhysRevLett.54.616} T.~Geisel, J.~Nierwetberg, and A.~Zacherl,
        \newblock {\em Phys. Rev. Lett.} \textbf{54}, 616 (1985).

    \bibitem{Barthelemy:2008rt} P.~Barthelemy, J.~Bertolotti, and D.~S. Wiersma,
        \newblock {\em Nature (London)} \textbf{453}, 495 (2008).

    \bibitem{barthelemy:2010PRE} P. Barthelemy, J. Bertolotti, K. Vynck, S.
        Lepri and D. S. Wiersma, Phys. Rev. E \textbf{82}, 011101 (2010).

    \bibitem{beenakker:2012} C.~W. Groth, A.~R. Akhmerov, and C.~W.~J.
        Beenakker, \newblock {\em Phys. Rev. E} \textbf{85}, 021138 (2012).

    \bibitem{svenson:2013PRE} T. Svensson, K. Vynck, M. Grisi, R. Savo, M.
        Burresi and D.S. Wiersma, Phys. Rev. E \textbf{87}, 022120 (2013).

    \bibitem{PhysRevE.61.1164} E.~Barkai, V.~Fleurov, and J.~Klafter, \newblock
        {\em Phys. Rev. E} \textbf{61}, 1164 (2000).

    \bibitem{Burioni:2010fk} R.~{Burioni}, L.~{Caniparoli}, and A.~{Vezzani},
        \newblock {\em Phys. Rev. E} \textbf{81}, 060101 (2010).

    \bibitem{Disanto} R.~{Burioni}, S.~{di Santo}, S.~{Lepri} and A.~{Vezzani},
        \newblock {\em Phys. Rev. E} \textbf{86}, 031125 (2012).

    \bibitem{Burioni:2010qy} R.~{Burioni}, L.~{Caniparoli}, S.~{Lepri}, and
        A.~{Vezzani}, \newblock {\em Phys. Rev. E} \textbf{81}, 011127 (2010).

    \bibitem{Beenakker:2009ys} C.W.J. Beenakker,  C.W. Groth and
        A.R. Akhmerov, \newblock {\em Phys. Rev. B} \textbf{79}, 024204
        (2009).

    \bibitem{Buonsante:2011} P.~{Buonsante}, R.~{Burioni}, and A.~{Vezzani},
        \newblock {\em Phys. Rev. E} \textbf{84}, 021105 (2011).

    \bibitem{Biazzo:2009qy} I.~{Biazzo}, F.~{Caltagirone}, G.~{Parisi}, and
        F.~{Zamponi}, \newblock{\em Phys. Rev. Lett.} \textbf{102}, 195701
        (2009).

    \bibitem{Parisi:2010} G. Parisi and  F. Zamponi, Rev. Mod. Phys. \textbf{82},
        789 (2010).

    \bibitem{Torquato:2010} S.~Torquato and  F. H. Stillinger, Rev. Mod. Phys.
        \textbf{82}, 3197 (2010).

    \bibitem{ann1} T. Geisel, J. Nierwetberg and A. Zacherl, Phys. Rev. Lett.
        \textbf{54}, 616 (1985).

    \bibitem{ann2} M.~F. Shlesinger, G.~M. Zaslavski and J.~Klafter, Nature
        (London) \textbf{363}, 31 (1993).

    \bibitem{ann3} G. Zumofen and J.~Klafter, Phys. Rev. E \textbf{47}, 851
        (1993).

    \bibitem{singler} T. Svensson, K. Vynck, E. Adolfsson, A. Farina, A. Pieri
        and D.S. Wiersma, {\it arXiv:1310.6419v1} (2013).

    \bibitem{0953-8984-16-37-002} T.~Okubo and T.~Odagaki, \newblock {\em Journal of Physics: Condensed Matter} \textbf{16}, 6651 (2004).

    \bibitem{Packing_Xu} N.~Xu, J.~Blawzdziewicz, and C.~S. O'Hern, \newblock {\em Phys. Rev. E} \textbf{71}, 061306 (2005).

    \bibitem{cates} M.E. Cates J. Phys. (Paris) \textbf{46}, 1059 (1985).
\end{thebibliography}

\end{document}